\documentclass[aps,prl,reprint,10pt,nofootinbib,superscriptaddress]{revtex4-2}

\normalsize
\usepackage{amsmath,amssymb,bm}
\usepackage{slashed}
\usepackage{graphicx}
\usepackage{booktabs}
\usepackage{placeins}
\usepackage{array}
\usepackage{microtype}
\usepackage{upgreek}
\usepackage{siunitx}
\usepackage{xcolor}
\usepackage{hyperref}
\hypersetup{hidelinks}

\newcommand{\supplementalmaterialref}[1]{%
  \hyperref[#1]{supplemental material~\ref*{#1}}%
}
\newcommand{\PiTwo}{\Pi_{2}}
\newcommand{\PiOne}{\Pi_{1}}
\newcommand{\cW}{c_{2W}}
\newcommand{\sWsq}{s_W^2}
\newcommand{\sW}{s_W}
\newcommand{\GF}{G_F}
\newcommand{\Kfac}{K}
\newcommand{\Lagr}{L}
\newcommand{\dd}{\mathrm d}
\newcommand{\Tr}{\mathrm{Tr}}
\newcommand{\sumint}{\mathop{\textstyle\sum\mkern-20mu\displaystyle\int}\nolimits}
\newcommand{\chemMu}{\upmu}
\makeatletter
\newenvironment{wideequations}{%
  \par\ignorespaces\onecolumngrid\vskip6\p@\prep@math@patch
}{%
  \par\vskip6\p@\twocolumngrid\global\@ignoretrue\@endpetrue
}
\makeatother

\begin{document}

\title{Comment on ``Long-range magnetic fields in the ground state of the Standard Model plasma'' }

\author{Jingwei Wang}
\author{Yiming Liu}
\email{liuym@cqu.edu.cn}
\author{Ligong Bian}
\email{lgbycl@cqu.edu.cn}
\affiliation{Department of Physics and Chongqing Key Laboratory for Strongly Coupled Physics, Chongqing University, Chongqing 401331, P. R. China}

\begin{abstract}

We revisit the two-loop parity-odd photon polarization analyzed by Boyarsky, Ruchayskiy, and Shaposhnikov [Phys.
Rev. Lett. 109, 111602 (2012)]. The original calculation neglected opposite-fermion-flow partner diagrams for the tadpole and exchange-bubble topologies alike. We quantify the impact of the properly restored exchange-bubble coefficients on the static magnetic response.

\end{abstract}

\maketitle

\emph{Introduction.---}
For a static electromagnetic field in an isotropic plasma, the parity-odd part of the photon polarization tensor can be written as
\begin{equation}
\Pi_{\rm odd}^{\mu\nu}(K_1)
 = i\epsilon^{\mu\nu\alpha\beta}u_\alpha K_{1\beta}
   \PiTwo(K_1^2),
\quad u^\mu=(1,\bm 0).
\label{eq:PiOdd}
\end{equation}
Throughout this work, $K_1^\mu=(0,\bm {k})$ denotes the static external photon four-momentum. The external three-momentum  $\bm p$ in Eq.~(2) of Ref.~\cite{boyarsky2012} is opposite to the momentum $\bm k$ used for our directly evaluated amplitudes: $\bm p=-\bm k$.  A nonzero long-wavelength limit $\PiTwo(0)$ generates a three-dimensional Chern--Simons term and can trigger a magnetic instability. Ref.~\cite{boyarsky2012} related this coefficient to the axial anomaly. For charged fermions with different left- and right-handed chemical potentials, the one-loop result has the form
\begin{equation}
\PiTwo(0)=\frac{\alpha}{2\pi}\,\Delta\chemMu,
\label{eq:oneloopCS}
\end{equation}
where the numerical coefficient is fixed by the anomalous axial Ward identity. Ref.~\cite{boyarsky2012} then argued that weak-interaction self-energy corrections shift the left- and right-handed dispersion relations by different amounts, $\delta\chemMu_L$ and $\delta\chemMu_R$, so that Eq.~\eqref{eq:oneloopCS} is used with $\Delta\chemMu\to \Delta\chemMu_{\rm tree}+\delta\chemMu$.

The issue addressed here is the two-loop amplitude and its complete fermion-flow content, rather than the renormalization of an axial Ward coefficient. The axial anomaly is one-loop exact and its coefficient receives no higher-loop renormalization, including at finite temperature and density Ref.~\cite{Adler:1969er,Itoyama:1982up,Contreras:1987ku,Qian:1994pp}. This non-renormalization theorem does not preclude genuine two-loop contributions to the finite-medium parity-odd photon polarization, since such contributions do not renormalize the anomaly coefficient. In particular, Ref.~\cite{boyarsky2012} omits the opposite-flow partners of its tadpole and exchange-bubble topologies. Ref.~\cite{dvornikov2014}, Eq.~(3), includes opposite-flow Wick contractions but assigns a cancellation to the exchange bubbles, whereas their two orientations add. We test these topological assertions by retaining both flow orientations and projecting the complete static amplitudes onto $i\epsilon_{ijk}k_k$. Refs.~\cite{boyarsky2012,notzold1988} show how the local parts of the two-loop tadpole and exchange-bubble self-energies can be incorporated by shifting the chiral fermion chemical potentials in the one-loop photon polarization, Fig.~1(a) of Ref.~\cite{boyarsky2012}. In the momentum-independent approximation, $\chemMu_{f}\to\chemMu_{f}+\delta\chemMu_{f}$ reproduces the corresponding self-energy insertion illustrated in its Fig.~1(b) of Ref.~\cite{boyarsky2012}. This treatment of the selected two-loop insertions is valid within that approximation; the discrepancy arises from the omitted opposite-fermion-flow topologies. 

%We test the tadpole diagrams pairing directly at the amplitude level and use diagrams 6 and 7 in Fig.~\ref{fig:selected17mma} as an explicit example of the same-sign exchange-bubble orientations. The remaining exchange-bubble diagrams are included separately in the coefficient comparison.
We adopt the same setup, conventions, and regularization scheme as Boyarsky, Ruchayskiy, and Shaposhnikov in Ref. ~\cite{boyarsky2012}.
In particular, we use the same finite-temperature fermion and photon propagators, parity-odd projection, and static limit prescription. No new modifications to the framework are introduced; we only restore the independent partner diagrams with opposite fermion flow, which were omitted in their diagrammatic analysis.

\begin{figure*}[!htp]
\centering
\includegraphics[width=0.9\textwidth]{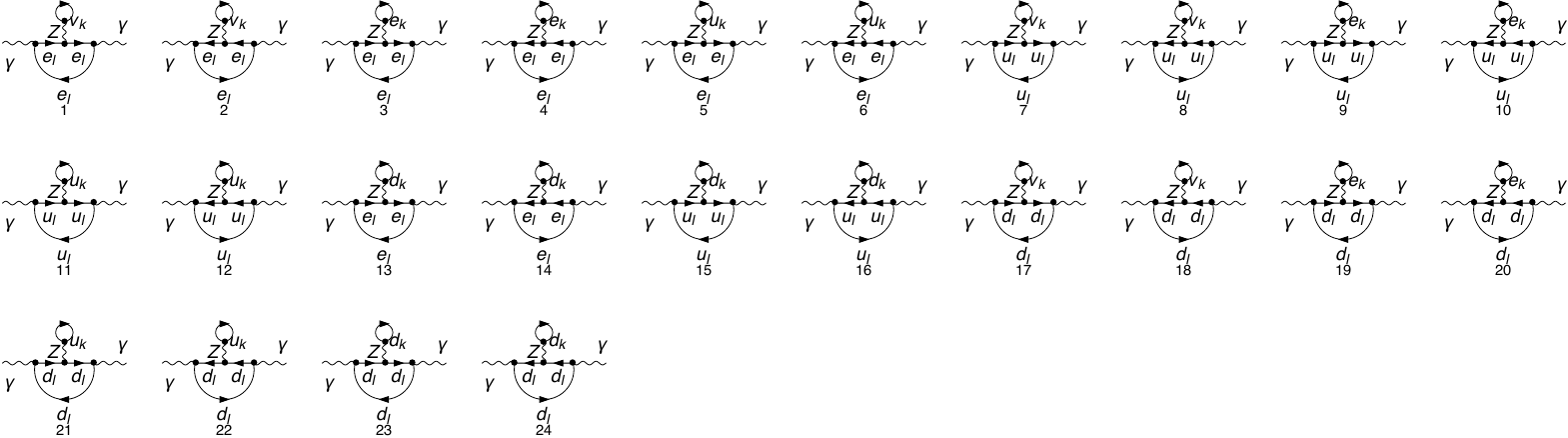}
\caption{Tadpole diagrams in the two-loop photon polarization.}
\label{fig:tadpoles}
\end{figure*}

\begin{figure*}[!htp]
\centering
\includegraphics[width=0.9\textwidth]{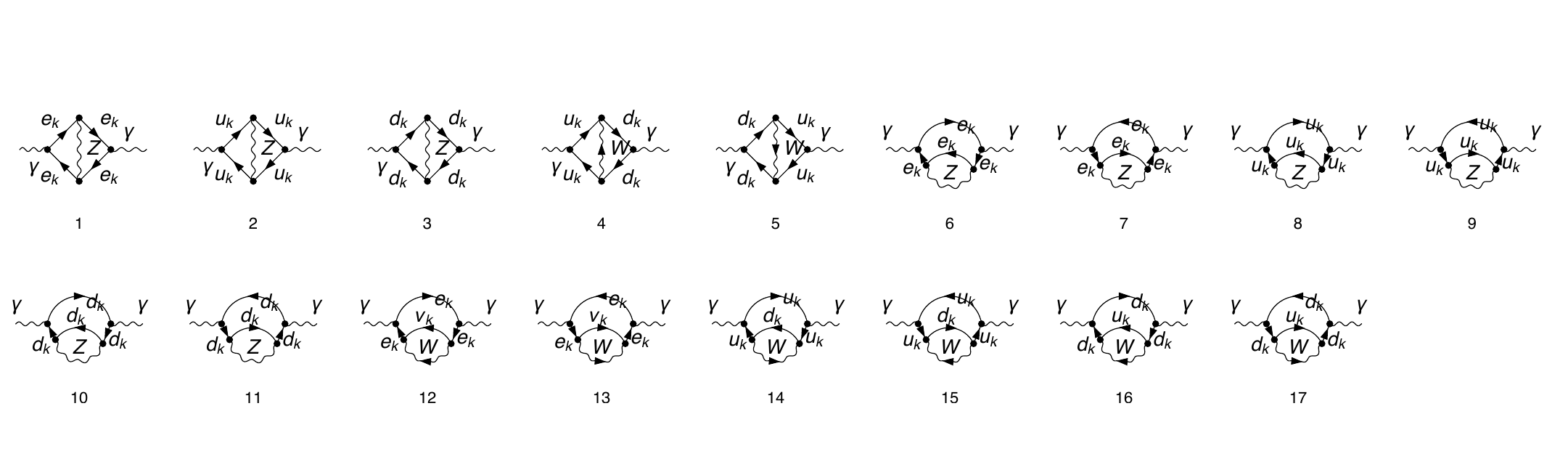}
\caption{Bubble and ladder diagrams in the two-loop photon polarization.}
\label{fig:selected17mma}
\end{figure*}

\emph{Low-energy theory and amplitude projection.---}
We work in the regime $|\chemMu_f|,\,|p_{\rm ext}|,\,|k_{\rm ext}|\ll T\ll M_W,M_Z$, where $p_{\rm ext}$ denotes a physical external fermion momentum and $k_{\rm ext}$ the external photon momentum. Here $T$ is the plasma temperature and $\chemMu_f$ are the microscopic chemical potentials. We use the four-fermion interaction of Ref.~\cite{boyarsky2012},
\begin{equation}
\Lagr_F=\frac{4\GF}{\sqrt2}
\left[(J_\rho^{\rm NC})^2+2J_\rho^{\rm CC,+}J^{\rm CC,-\rho}\right],
\label{eq:FermiLag}
\end{equation}
with
\begin{align}
J_\rho^{\rm CC,+}&=\frac{1}{\sqrt2}
\left(\bar\ell_L\gamma_\rho\nu_L+\bar d_L\gamma_\rho u_L+\bar s_L\gamma_\rho c_L+\ldots\right),\\
J_\rho^{\rm NC}&=\sum_\psi \bar\psi\gamma_\rho
\left(g_L^\psi P_L+g_R^\psi P_R\right)\psi.
\end{align}
Here, $g_L^\psi=T_3^\psi-\sWsq Q_\psi$, $g_R^\psi=-\sWsq Q_\psi$ and
$\sW= \sin \theta_W$ are the left-handed coupling, the right-handed coupling, and the sine of the weak mixing angle, respectively.
The five-flavor calculation excludes the $Zt\bar t$ and $Wtb$ vertices because no top-quark lines are included among its thermally active fermion species; both vertices exist in the full Standard Model.
The total topologies under study include 24 tadpole diagrams and the 17 non-tadpole diagrams, as displayed in Figs.~\ref{fig:tadpoles} and \ref{fig:selected17mma}, respectively. 

%\emph{Two-loop amplitudes and fermion-flow orientations.---}
%\label{sec:bubble}
The photon polarization is evaluated directly at two loops in the four-fermion effective theory of Eq.~\eqref{eq:FermiLag}. We generate the photon two-point topologies with two electromagnetic vertices and one weak four-fermion interaction, evaluate their Dirac traces with chiral projectors, and perform the Matsubara sums and dimensionally regularized spatial momentum integrations. The parity-odd coefficient is obtained by projecting the static amplitude onto the term linear in the external spatial photon momentum and the fermion chemical potentials. Both fermion-flow orientations are included, and the three lepton flavors are summed only after the individual contributions have been identified.

\emph{Cancellation contributions from tadpole diagrams.---}
\label{sec:tadpoleTrace}
We find all the tadpole diagrams' summation contribute zero Chern-Simons term, due to the cancellation of opposite-fermion-flow tadpole amplitudes.  We use FeynArts~\cite{kublbeck1990} to generate Feynman diagrams and FeynCalc~\cite{mertig1991,shtabovenko2016} to obtain the corresponding amplitudes. We then extract the part proportional to the rank-three antisymmetric tensor and expand the amplitudes to first order in both the fermion chemical potential and the external photon momentum $k_1$, retaining the $O(k_1\chemMu$) terms. For the finite temperature calculations, the fermionic Matsubara sum and spatial momentum integral follow the finite-temperature convention of Ref.~\cite{sirena2026}. In the following, we demonstrate that with the first two diagrams in Fig.~\ref{fig:tadpoles}. The two opposite-fermion-flow tadpole amplitudes in Fig.~\ref{fig:tadpoles}, diagrams 1 and 2, follow directly from the unexpanded fermion chains. For convenience in the following discussion, let us set $P_{R,L}=(1\pm\gamma^5)/2$, and define the effective Z-fermion neutral-current vertex factors for the two fermion-flow orientations: 
\begin{equation}
\begin{aligned}
\Gamma_{-,\rho}&=-ie\gamma_\rho\left(\frac{s_W}{c_W}P_L+\frac{s_W^2-\frac12}{c_Ws_W}P_R\right),\\
\Gamma_{+,\rho}&=ie\gamma_\rho\left(\frac{s_W}{c_W}P_R+\frac{s_W^2-\frac12}{c_Ws_W}P_L\right).
\end{aligned}
\label{eq:tadpoleVertices}
\end{equation}
To write the two amplitudes compactly, we introduce the neutrino-loop factor, the common photon-chain numerator, the two trace factors, and the common denominator: 
\begin{align*}
    J_R^\rho&=\Tr\!\left[\frac{ie}{2c_Ws_W}
    \slashed L_1\gamma^\rho P_R\right],\\
 Y_{\rm tad}^{\mu\nu}
 &=e^2(-\slashed L_2)\gamma^\nu
   (\slashed L_2+\slashed K_1)\gamma^\mu\slashed L_2,\\
 T_{{\rm tad},\rho,1}^{\mu\nu}=\Tr [&Y_{\rm tad}^{\mu\nu}\Gamma_{-,\rho}],\qquad T_{{\rm tad},\rho,2}^{\mu\nu}=\Tr [Y_{\rm tad}^{\mu\nu}\Gamma_{+,\rho}],\\
 D_{\rm tad}&=L_1^2(L_2^2)^2(L_2+K_1)^2(-M_Z^2),
\end{align*}
respectively. Here, integrating out the heavy neutral weak boson at low temperature gives rise to
 $D_Z^{\mu\nu}(q)
=i g^{\mu\nu}/M_Z^2
+\mathrm O\!\left(q^2/M_Z^4\right)$. Then the two amplitudes obtained before expanding in external momentum or chemical potential are
\begin{equation}
\begin{aligned}
&\Pi_{{\rm tad},j}^{\mu\nu}(K_1)=i\sum_{g}
 \sumint_{L_1,L_2}\!\frac{J_R^\rho T_{{\rm tad},\rho,j}^{\mu\nu}}{D_{\rm tad}},\quad j=1,2,
\end{aligned}
\label{eq:tadpoleRawPair}
\end{equation}
where $\sum\limits_g$ denotes the sum over the three generations, and the definition of the Matsubara integral is given in supplemental material. It should be noted that, to unify the final form, we assign a consistent momentum direction to the internal momentum $L_2$ in Fig.~\ref{fig:tadpoles}, rather than following the directions of the respective arrows.

Following the finite-density weak self-energy treatment of Refs.~\cite{notzold1988,boyarsky2012}, the term linear in the fermion chemical potential is inserted only in the separate tadpole fermion loop with momentum $L_1$; the photon-carrying fermion loop with momentum $L_2$ is evaluated at zero chemical potential.
For convenience of further discussion, we define charge conjugation as follows: 
\begin{align*}
     \widehat X&= C^{-1}X^TC,
\end{align*}
where $C$ is the charge-conjugation matrix. 
With
$\widehat{\gamma^\alpha}=-\gamma^\alpha$, $\widehat{\gamma^5}=\gamma^5$ and $\widehat{AB}=\widehat{B}\widehat{A}$
, the reversed chain obeys
\begin{equation}
\begin{aligned}
 T_{{\rm tad},1}^{\mu\nu}
 &=\Tr\!\left[\widehat{\Gamma_{-,\rho}}\widehat{Y_{\rm tad}^{\mu\nu}}\right],\\
 \widehat{\Gamma_{-,\rho}}&=\Gamma_{+,\rho},\\
 \widehat{Y_{\rm tad}^{\mu\nu}}(L_2,K_1)
 &=Y_{\rm tad}^{\nu\mu}(-L_2,-K_1),\\
 \implies T_{{\rm tad},1}^{\mu\nu}(L_2,K_1)
 &=T_{{\rm tad},2}^{\nu\mu}(-L_2,-K_1).
\end{aligned}
\label{eq:tadpoleChargeConjugation}
\end{equation}
At $O(k_1\chemMu)$, the parity-odd tadpole trace is proportional to the rank-three antisymmetric tensor $i\epsilon^{ijk}k\chemMu$. In the reversed-index comparison, the $\gamma^5$ term of $T_{{\rm tad},2}^{\nu\mu}$ has the opposite sign to that of $T_{{\rm tad},1}^{\mu\nu}$. The paired tadpole contribution thus vanishes in the static projection.

\emph{$Zero$ contribution of Ladder diagrams.---}
\label{sec:ladderTrace}
The Ladder diagrams contribution is zero due to Fierz
factorization as will be shown bellow. 

For the Feynman diagram 1 in Fig.~\ref{fig:selected17mma}, we introduce
\begin{equation}
\begin{aligned}
 X^\nu&=(\slashed{K}_1-\slashed{L}_2)\gamma^\nu(-\slashed{L}_2),\\
 Y^\mu&=\slashed{L}_1\gamma^\mu(\slashed{K}_1+\slashed{L}_1),\\
 G&=g_L^\psi P_L+g_R^\psi P_R.
\end{aligned}
\label{eq:ladderInputMain}
\end{equation}
Since $[Y^\mu\gamma_\rho,\gamma^5]=0$, the mixed-chirality terms vanish. Applying the strictly four-dimensional Fierz identity
\begin{equation}
 (\gamma^\rho P_\chi)_{ab}(\gamma_\rho P_\chi)_{cd}
 =-(\gamma^\rho P_\chi)_{ad}(\gamma_\rho P_\chi)_{cb}
\label{eq:ladderFierzMain}
\end{equation}
gives the exact rearrangement of the original single trace,
\begin{equation}
\begin{aligned}
 &\Tr[X^\nu\gamma^\rho GY^\mu\gamma_\rho G]\\
 &\quad=-\sum_{\chi=L,R}g_\chi^2
  \Tr[X^\nu\gamma^\rho P_\chi]\,
  \Tr[Y^\mu\gamma_\rho P_\chi].
\end{aligned}
\label{eq:ladderFactorizedMain}
\end{equation}
Each factorized photon-current correlator begins at first order in the static external momentum. Hence
\begin{equation}
 \Pi_{\rm lad}^{\mu\nu}(k)=O(|\bm k|^2),\qquad
 \left.\Pi_{2}^{\rm lad}(0)\right|_{O(|\bm k|)}=0.
\label{eq:ladderNoLinearMain}
\end{equation}
This argument applies to the ladder topology represented by diagrams 1--5 of Fig.~\ref{fig:selected17mma} in the local weak-boson limit.

\emph{Double contributions from exchange bubble diagrams.---}
The calculations of exchange bubble diagrams in Ref.~\cite{boyarsky2012} overlooks the opposite-fermion-flow partners. And, Ref.~\cite{dvornikov2014} assigned the wrong cancellation in these diagrams. In the following, we select Feynman diagrams 6 and 7 of Fig.~\ref{fig:selected17mma} as an explicit example to demonstrate the issue. 

The first term of Eq.~(7) in Ref.~\cite{boyarsky2012} is the same-flavor, left-handed exchange insertion. For a charged lepton $\ell$, with $g_L^\ell=-\tfrac12+s_W^2$, its contribution to the anomaly-mediated coefficient is
\begin{equation}
 \left.\Pi_{2,\ell}^{\rm (7),ex}(0)\right|_{L}
 =\frac{K T^2}{3}(g_L^\ell)^2\chemMu_{\ell L}
 =\frac{KT^2}{12}(1-2s_W^2)^2\chemMu_{\ell L},
 \label{eq:Eq7FirstBubble}
\end{equation}
where $\Kfac= \frac{\alpha}{2\pi}\frac{4\GF}{\sqrt2}$.
Including the right-handed counterpart with $g_R^\ell=s_W^2$ and taking the left--right difference gives
\begin{align}
\Pi_{2,\ell}^{\rm (7),ex}(0)
=\frac{KT^2}{12}X,
 \label{eq:Eq7BubblePair}
\end{align}
where $X=(1-2s_W^2)^2\chemMu_{\ell L} -4s_W^4\chemMu_{\ell R}$.
The direct, strictly two-loop evaluation gives equal parity-odd amplitudes for the two opposite fermion-flow orientations, diagrams 6 and 7 of Fig.~\ref{fig:selected17mma}. In the external-momentum convention $k$ of the two-loop diagrams,
\begin{equation}
 \Pi^{\mu\nu}_{6,\rm 2L}=
 \Pi^{\mu\nu}_{7,\rm 2L}
 =-\frac{e^4T^2}{48\pi^2M_Z^2s_W^2c_W^2}
 E^{\mu\nu}(k)X,
 \label{eq:Direct67Comparison}
\end{equation}
where $E^{\mu\nu}(k)=\epsilon^{\mu\nu\rho}k_\rho$. The evaluated amplitudes include a sum over three charged-lepton flavors. To compare with the single-flavor term in Eq.~\eqref{eq:Eq7BubblePair}, first divide by three. Next use the opposite external-momentum routing of the reference, $p=-k$, so that $E^{\mu\nu}(k)=-E^{\mu\nu}(p)$. Defining $B_{i,\rm ref}^{(1\ell)}$ as the single-flavor scalar coefficient extracted against $E^{\mu\nu}(p)$, rather than $E^{\mu\nu}(k)$, yields
\begin{align}
 B_{6,\rm ref}^{(1\ell)}
 &=B_{7,\rm ref}^{(1\ell)}=\frac{KT^2}{9}X,
 \\
 \frac{B_{6,\rm ref}^{(1\ell)}}
 {\Pi_{2,\ell}^{\rm (7),ex}(0)}
 &=\frac43.
 \label{eq:bubbleRatio}
\end{align}
 For the two opposite fermion-flow orientations of Fig.~\ref{fig:selected17mma}, following the weak self-energy treatment of Refs.~\cite{notzold1988,boyarsky2012}, only the fermion propagator with momentum $L_2$ in the $Z$-exchange subloop carries the linear chemical-potential insertion; the other fermion propagators, with momenta $L_1$ and $Q=L_1-K_1$, retain their zero-chemical-potential forms. Thus, similarly, we define the denominators for the two fermion-flow orientations, common photon-chain factor, and the numerator trace, respectively, as
\begin{align*}
    D_{\rm src,j}&=L_1^2 L_{2,j}^2 (Q^2)^2(-M_Z^2),\\
    Y^{\mu\nu}(Q,L_1)
 &=e^2(-\slashed Q)\gamma^\nu
 \slashed L_1\gamma^\mu\slashed Q,\\
  T_6^{\mu\nu}&=\Tr [Y^{\mu\nu}(Q,L_1)\Gamma_+^\rho
 \slashed L_{2,6}\Gamma_{+,\rho}],\\
 T_7^{\mu\nu}&=\Tr [Y^{\mu\nu}(Q,L_1)\Gamma_-^\rho
 \slashed L_{2,7}\Gamma_{-,\rho}],
\end{align*}
where $Q=L_1-K_1$, $\slashed L_{2,j}=\slashed L_{2,0}
 +\chemMu_j\gamma^0$ for $j=6,7$ and $L_{2,0}=(i(2n+1)\pi T,\bm\ell_2)=(i\omega_n,\bm\ell_2)$ . Then the unexpanded Minkowski amplitudes are
\begin{gather}
\begin{aligned}
 \Pi_{j}^{\mu\nu}(K_1)
 =-i\sum_{g}
  \sumint_{L_1,L_2}\!\frac{T^{\mu\nu}_j}{D_{\rm src,j}},\qquad j=6,7
\end{aligned}
\label{eq:bubbleSource12}
\end{gather}
Transposing the complete ordered matrix product gives
\begin{equation*}
\begin{aligned}
 T_6^{\mu\nu}
 &=\Tr [Y^{\mu\nu}\Gamma_+^\rho
 \slashed L_{2,6}\Gamma_{+,\rho}]\\
 &=\Tr [\widehat\Gamma_{+,\rho}\,
 \widehat{\slashed L}_{2,6}\,
 \widehat\Gamma_+^\rho\,
 \widehat Y^{\mu\nu}].
\end{aligned}
\end{equation*}
The photon chain transforms as
\begin{equation*}
\begin{aligned}
 \widehat Y^{\mu\nu}(Q,L_1)
 &=e^2(-\slashed Q)(-\gamma^\mu)
 (-\slashed L_1)(-\gamma^\nu)\slashed Q\\
 &=e^2\slashed Q\gamma^\mu
 \slashed L_1\gamma^\nu\slashed Q\\
 &=Y^{\nu\mu}(-Q,-L_1).
\end{aligned}
\end{equation*}
The reversal also determines the directed momenta:
\begin{equation*}
\begin{aligned}
 L_1'&=-L_1,\quad K_1'=-K_1,\quad L_2'=-L_2,\\
 Q'&=L_1'-K_1'=-Q\;.
\end{aligned}
\end{equation*}
For the fermion numerator, $\slashed L_{2,j}(P)\equiv\slashed P+\chemMu_j\gamma^0$; the parentheses indicate the momentum argument, not multiplication. The chemical-potential assignment then follows from
\begin{equation*}
\begin{aligned}
 \widehat{\slashed L}_{2,6}(L_{2,0})
 &=\slashed L_{2,7}(-L_{2,0})
\Longrightarrow \chemMu_7=-\chemMu_6=-\chemMu_e.
\end{aligned}
\end{equation*}
The corresponding denominators obey
\begin{equation*}
\begin{aligned}
 L_{2,7}^2(-\omega_{n},-\bm\ell_2;\chemMu_7)
 &=L_{2,6}^2(\omega_{n},\bm\ell_2;\chemMu_6)\\
 \implies D_{\rm src,6}&=D_{\rm src,7}(L_1',L_2',K_1';\chemMu_7)\;.
\end{aligned}
\end{equation*}
Finally, the numerator  $T_6^{\mu\nu}$ is transformed into $T_7^{\nu\mu}$ by reordering the matrices inside the trace :
\begin{equation*}
\begin{aligned}
 T_6^{\mu\nu}
 &=\Tr\big[Y^{\nu\mu}(Q',L_1')\,
 \Gamma_-^\rho\,
 \slashed L_{2,7}(L_{2,0}')\,\Gamma_{-,\rho}\big]\\
 &=T_7^{\nu\mu}(L_1',L_2',K_1';\chemMu_7).
\end{aligned}
\end{equation*}
Their parity-odd traces are proportional to the external momentum $k_1$ and the fermion chemical potential. The $\gamma^5$ term in $T_7^{\nu\mu}$ has the opposite sign to that in $T_6^{\mu\nu}$; with the charge-conjugate assignment $\chemMu_7=-\chemMu_6$ and the common momentum routing established above, their contributions at $O(k_1\chemMu)$ are equal.

The factor $4/3$ is attributed here to the different treatment of the fermion chemical-potential corrections: the present work retains the strict two-loop contribution, whereas Ref.~\cite{boyarsky2012} uses the self-energy-corrected propagator of its Eq.~(8), which incorporates chemical-potential corrections beyond a single weak self-energy insertion. Thus the two treatments differ in whether higher-loop corrections to the fermionic chemical potentials are included. 

The contributions from the two bubble diagrams add, resulting in an overall factor of two. However, Ref.~\cite{boyarsky2012} considered only one of the two topologies, while the Ref.~\cite{dvornikov2014} incorrectly concluded that the contributions from these two topologically distinct diagrams, which have different fermion-flow directions, cancel each other.

\emph{Comparison with the complete coefficient of Ref.~\cite{boyarsky2012}.---}
Figure~\ref{fig:tadpoles} shows the tadpole diagrams. Their static linear tadpole projection vanishes after the pairing in Eq.~\eqref{eq:tadpoleRawPair}. Ref.~\cite{boyarsky2012} presents tadpole and exchange contributions together in Eq.~(B1). The coefficient tables below retain the original, separately computed exchange-bubble contribution of diagrams 6--17 in Fig.~\ref{fig:selected17mma}.

For relativistic fermions, to first order in the microscopic chemical potentials,
\begin{equation}
\Delta n_\nu=\frac{T^2}{6}\chemMu_\nu,
\qquad \Delta n_\ell=\frac{T^2}{3}\chemMu_\ell,
\qquad \Delta n_q=T^2\chemMu_q.
\label{eq:deltan}
\end{equation}
In the five-quark-flavor approximation, Eq.~(B1) of Ref.~\cite{boyarsky2012} is
\begin{align}
\frac{\Pi_{2}^{\rm (B1)}(0)}{K}
={}&-\frac29\sum_\alpha\Delta n_{\nu_\alpha}
-\frac{31}{36}(1-2\cW)\sum_\alpha\Delta n_{\ell_\alpha}
\nonumber\\
&+\frac{17-62\cW}{81}(\Delta n_u+\Delta n_c)
\nonumber\\
&+\frac{91+134\cW}{324}(\Delta n_d+\Delta n_s)
\nonumber\\
&+\frac{67}{324}(1+2\cW)\Delta n_b.
\label{eq:refB1full}
\end{align}
Equation~\eqref{eq:deltan} turns the reference expression into the ``Ref. total'' coefficients displayed in Table~\ref{tab:analytic}. The directly evaluated diagrams 6--17 are compared with the reference in Table~\ref{tab:analytic}.

To express the results in the conserved-charge basis of Ref.~\cite{boyarsky2012}, let $b$, $q$, and $\ell_\alpha$ be the thermodynamic multipliers conjugate to baryon number, electric charge, and lepton flavor. With five active quarks,
\begin{align}
\chemMu_{\nu_\alpha}&=\ell_\alpha,
&\chemMu_{\ell_\alpha}&=\ell_\alpha-q,
\nonumber\\
\chemMu_u=\chemMu_c&=\frac b3+\frac{2q}{3},
&\chemMu_d=\chemMu_s=\chemMu_b&=\frac b3-\frac q3.
\label{eq:mumultipliers}
\end{align}
Electric neutrality and the density relations in Eq.~\eqref{eq:deltan} give, with $L=\sum_\alpha L_\alpha$,
\begin{equation}
\sum_\alpha\chemMu_{\nu_\alpha}=\frac{-6B+66L}{23T^2},
\quad
\sum_\alpha\chemMu_{\ell_\alpha}=\frac{3B+36L}{23T^2},
\label{eq:musums}
\end{equation}
\begin{equation}
\begin{aligned}
\chemMu_{u,c}=\frac{6(2B+L)}{23T^2},\quad
\chemMu_{d,sb}=\frac{15B-4L}{23T^2}.
\end{aligned}
\label{eq:muquarks}
\end{equation}
The full coefficient of the reference, obtained by substituting these relations into Eq.~\eqref{eq:refB1full}, is
\begin{align}
\frac{\Pi_{2}^{\rm (B1)}(0)}{K}
&=c_L^{\rm ref}L+c_B^{\rm ref}B,
\label{eq:refchargeform}\\
\left(c_L^{\rm ref},c_B^{\rm ref}\right)
&=\left(\frac{8(22\cW-45)}{621},
\frac{53\cW+430}{621}\right).
\label{eq:refclexact}
\end{align}

For the separately computed exchange-bubble response of diagrams 6--17 in Fig.~\ref{fig:selected17mma}, write $\Pi_{2}^{\rm ins}$. It gives
\begin{align}
\frac{\Pi_{2}^{\rm ins}(0)}{K}
&=c'_L L+c'_B B,
\label{eq:thischargeform}\\
\left(c'_L,c'_B\right)
&=\left(\frac{8(68\cW+57)}{621},
\frac{4(88\cW+113)}{621}\right).
\label{eq:thisclexact}
\end{align}
With $\sin^2\theta_W=0.23122$, the comparison between the expression of Ref.~\cite{boyarsky2012} and the exchange-bubble diagrams 6--17 of Fig.~\ref{fig:selected17mma} is
\begin{align}
\left(c'_L,c'_B\right)
&=(1.20521,\;1.03256),\nonumber\\
\left(c_L^{\rm ref},c_B^{\rm ref}\right)
&=(-0.42736,\;0.73831).
\label{eq:chargenumeric}
\end{align}

\emph{Conclusion and discussion.---}
The tadpole diagrams in Fig.~\ref{fig:tadpoles} cancel. The Fierz decomposition in Eq.~\eqref{eq:ladderFactorizedMain} shows that the ladder amplitude from diagrams 1 to 5 in the Fig.~\ref{fig:selected17mma} has no term linear in the external photon momentum. In contrast, the rest paired exchange-bubble diagrams of 6-17 in Fig.~\ref{fig:selected17mma} have equal parity-odd amplitudes after reversing the fermion flow.

From Eqs.~\eqref{eq:refchargeform} and \eqref{eq:thischargeform}, the full coefficient quoted in Ref.~\cite{boyarsky2012} and the separately evaluated exchange-bubble coefficient of diagrams 6--17 are, 
\begin{align}
 \Pi_2^{\rm ref}(0)&=K(-0.42736L+0.73831B),\nonumber\\
 \Pi_2^{\rm ins}(0)&=K(1.20521L+1.03256B).
 \label{eq:implicationDelta}
\end{align}
These are a comparison of distinct sectors: the reference expression includes the tadpole term, while $\Pi_2^{\rm ins}$ contains only exchange-bubble diagrams 6--17. The following consequences therefore apply conditionally if $\Pi_2^{\rm ins}$ is used in place of the complete coefficient, with the remaining plasma parameters held fixed.

Define the signed coefficient ratio, wherever $\Pi_2^{\rm ref}(0)\ne0$, by
\begin{equation}
 R(L,B)=\frac{\Pi_2^{\rm ins}(0)}{\Pi_2^{\rm ref}(0)}
 \label{eq:implicationGeneralRatio}
\end{equation}
Equation~\eqref{eq:implicationGeneralRatio} determines the change in the characteristic unstable wave number through $|\Pi_2^{\rm ins}(0)|=|R(L,B)|\,|\Pi_2^{\rm ref}(0)|$. For otherwise identical plasma conditions, $|R|>1$ increases the linear growth exponent and $|R|<1$ decreases it.

The parity-odd coefficient enters the static transverse photon kernel as
\begin{equation}
 \Pi_{ij}(\bm p)
 =(p^2\delta_{ij}-p_ip_j)\PiOne(p^2)
 +i\epsilon_{ijk}p_k\PiTwo(p^2).
 \label{eq:conclusionPi2}
\end{equation}
For $\bm p=p\hat{\bm z}$, the transverse block obeys
\begin{equation}
 \Pi_T=\begin{pmatrix}
 p^2\PiOne & ip\PiTwo\\
 -ip\PiTwo & p^2\PiOne
 \end{pmatrix},
 \quad
 \det(\Pi_T-\lambda I)=0,
 \label{eq:implicationKernel}
\end{equation}
For the circular polarizations $\bm e_h=(1,ih,0)^T/\sqrt{2}$, with the helicity $h=\pm1$, the eigenvalue equation reads
\begin{equation}
 \begin{aligned}
 i\bm p\times\bm e_h&=hp\bm e_h,\\
 \Pi_T\bm e_h&=\lambda_h(p)\bm e_h,\\
 \lambda_h(p)&=p^2\PiOne(p^2)-hp\PiTwo(p^2).
 \end{aligned}
 \label{eq:helicityEigenvalues}
\end{equation}
The parity-even contribution is independent of $h$, whereas the Chern--Simons term distinguishes the two helicities. When $\PiOne(0)>0$, the of the helicity $h=\operatorname{sgn}[\PiTwo(0)]$ is negative for
\begin{equation}
 0<p<p_{\rm crit},\qquad
 p_{\rm crit}=\frac{|\PiTwo(0)|}{\PiOne(0)}.
 \label{eq:instabilityBand}
\end{equation}
For a conducting medium with conductivity $\sigma$, the linear growth rate in the long-wavelength approximation Ref.~\cite{Joyce:1997uy} is
\begin{equation}
 \gamma(p)=\frac{p|\PiTwo(0)|-p^2\PiOne(0)}{\sigma}.
 \label{eq:implicationGamma}
\end{equation}
The stationary condition gives
\begin{equation}
 \frac{\partial\gamma}{\partial p}
 =\frac{|\PiTwo(0)|-2p\PiOne(0)}{\sigma}=0,
 \qquad
 p_{\rm max}=\frac{|\PiTwo(0)|}{2\PiOne(0)}.
 \label{eq:implicationStationary}
\end{equation}
Evaluating $\gamma(p)$ at $p_{\rm max}$ yields
\begin{equation}
 \begin{aligned}
 \gamma_{\rm max}
 &=\frac{p_{\rm max}|\PiTwo(0)|
         -p_{\rm max}^{2}\PiOne(0)}{\sigma}=\frac{\PiTwo(0)^2}{4\sigma\PiOne(0)}.
 \end{aligned}
 \label{eq:implicationGrowth}
\end{equation}
Following Ref.~\cite{boyarsky2012}, use the characteristic-scale magnetic growth exponent $\beta_X=k_X^2t/\sigma$, with $B_X(t)/B_X(0)\sim e^{\beta_X}$. The wave number is estimated as $k_X=\xi|\Pi_2^X(0)|$, where the same order-one factor $\xi$ selects a mode in the unstable band for both coefficients. This is the growth estimation of Ref.~\cite{boyarsky2012}, not the exact maximized exponent $\gamma_{\rm max}t$ above. For equal $t$, $\sigma$, and $\xi$,
\begin{equation}
 \begin{aligned}
 \beta_X(t)&=\frac{k_X^2t}{\sigma},~
 \frac{\beta_{\rm ins}}{\beta_{\rm ref}}
 &=\left(\frac{k_{\rm ins}}{k_{\rm ref}}\right)^2=|R(L,B)|^2.
 \end{aligned}
 \label{eq:implicationBetaRatio}
\end{equation}
The charge backgrounds are summarized bellow:
\begin{center}
\small
\begin{tabular}{lcc}
%\label{tab:CB}
\toprule
Charge background & $|R|$ & $\beta_{\rm ins}/\beta_{\rm ref}$\\
\midrule
$B=0$ & 2.82 & 7.95\\
$L=0$ & 1.40 & 1.96\\
$L=-B$ ($B+L=0$) & 0.148 & 0.0219\\
\bottomrule
\end{tabular}
\end{center}
For the charge backgrounds in the table, the estimated linear growth of primordial magnetic fields is enhanced relative to the reference case for $B=0$ and $L=0$, whereas it is suppressed for $L=-B$.

Relative to Ref.~\cite{boyarsky2012}, the exchange-bubble coefficients enhance the estimated linear growth of primordial magnetic fields for a purely leptonic asymmetry ($B=0$) and for a purely baryonic asymmetry ($L=0$). For $L=-B$ with $B\ne0$, the parity-odd coefficient remains nonzero despite $B+L=0$. This charge background therefore does not exclude the electromagnetic Chern--Simons term of Ref.~\cite{boyarsky2012}.

In the conditional comparison above, $|\Pi_2^{\rm ins}(0)|\simeq0.148|\Pi_2^{\rm ref}(0)|$. The latter suppresses the estimated linear growth exponent, not the magnetic-field amplitude directly. Moreover, for \(L=-B\), \(R(-B,B)\simeq-0.148<0\), indicating that the Chern–Simons coefficient reverses sign relative to Ref. [1]. From Eq. (33), the circular-polarization eigenvalues are
\[
\lambda_h(p)=p^2\Pi_1(p^2)-hp\Pi_2(p^2),
\qquad h=\pm1.
\]
For \(\Pi_1(0)>0\), the unstable mode satisfies \(\lambda_h(p)<0\) in the momentum interval given by Eq. (34), which requires
\[
h=\operatorname{sgn}[\Pi_2(0)].
\]
Consequently, the sign reversal of the Chern–Simons coefficient implies
\[
h_{\mathrm{ins}}=-h_{\mathrm{ref}},
\]
showing that the two results favor opposite magnetic helicities under the same polarization convention.

\emph{Acknowledgments.---}
This work is supported by the National Natural Science Foundation of China (NSFC) under Grants Nos. 12675130, 12322505, and 12547101.
L.B. also acknowledges Chongqing Natural Science Foundation under Grant
No. CSTB2024NSCQ-JQX0022.  We thank Jie Liu and Haolin Li for helpful discussions.

\bibliographystyle{apsrev4-2}
\bibliography{references}

@article{Joyce:1997uy,
    author = "Joyce, M. and Shaposhnikov, Mikhail E.",
    title = "{Primordial magnetic fields, right-handed electrons, and the Abelian anomaly}",
    eprint = "astro-ph/9703005",
    archivePrefix = "arXiv",
    reportNumber = "CERN-TH-97-31",
    doi = "10.1103/PhysRevLett.79.1193",
    journal = "Phys. Rev. Lett.",
    volume = "79",
    pages = "1193--1196",
    year = "1997"
}

@article{Adler:1969er,
    author = "Adler, Stephen L. and Bardeen, William A.",
    title = "{Absence of higher order corrections in the anomalous axial vector divergence equation}",
    doi = "10.1103/PhysRev.182.1517",
    journal = "Phys. Rev.",
    volume = "182",
    pages = "1517--1536",
    year = "1969"
}

@article{Itoyama:1982up,
    author = "Itoyama, Hiroshi and Mueller, Alfred H.",
    title = "{The Axial Anomaly at Finite Temperature}",
    reportNumber = "CU-TP-256",
    doi = "10.1016/0550-3213(83)90370-X",
    journal = "Nucl. Phys. B",
    volume = "218",
    pages = "349--365",
    year = "1983"
}

@article{Contreras:1987ku,
    author = "Contreras, C. and Loewe, M.",
    title = "{The Triangle Diagram at Finite Temperature}",
    reportNumber = "Print-87-0836 (CHILE U.CATOLICA)",
    doi = "10.1007/BF01555887",
    journal = "Z. Phys. C",
    volume = "40",
    pages = "253--258",
    year = "1988"
}

@article{Qian:1994pp,
    author = "Qian, Zhi-Xin and Su, Ru-Keng and Yu, P. K. N.",
    title = "{Axial anomaly at finite temperature and finite density}",
    doi = "10.1007/BF01557631",
    journal = "Z. Phys. C",
    volume = "63",
    pages = "651--654",
    year = "1994"
}

@article{notzold1988,
  author  = {N{\"o}tzold, Dirk and Raffelt, Georg},
  title   = {Neutrino dispersion at finite temperature and density},
  journal = {Nucl. Phys. B},
  volume  = {307},
  pages   = {924--936},
  year    = {1988},
  doi     = {10.1016/0550-3213(88)90113-7}
}

@article{boyarsky2012,
    author = "Boyarsky, Alexey and Ruchayskiy, Oleg and Shaposhnikov, Mikhail",
    title = "{Long-range magnetic fields in the ground state of the Standard Model plasma}",
    eprint = "1204.3604",
    archivePrefix = "arXiv",
    primaryClass = "hep-ph",
    reportNumber = "CERN-PH-TH-2012-112",
    doi = "10.1103/PhysRevLett.109.111602",
    journal = "Phys. Rev. Lett.",
    volume = "109",
    pages = "111602",
    year = "2012"
}

@article{kublbeck1990,
  author  = {K{\"u}blbeck, J. and B{\"o}hm, M. and Denner, A.},
  title   = {Feyn Arts: Computer Algebraic Generation of Feynman Graphs and Amplitudes},
  journal = {Comput. Phys. Commun.},
  volume  = {60},
  pages   = {165--180},
  year    = {1990}
}

@article{mertig1991,
  author  = {Mertig, R. and B{\"o}hm, M. and Denner, A.},
  title   = {FEYN CALC: Computer algebraic calculation of Feynman amplitudes},
  journal = {Comput. Phys. Commun.},
  volume  = {64},
  pages   = {345--359},
  year    = {1991}
}

@article{shtabovenko2016,
  author        = {Shtabovenko, Vladyslav},
  title         = {FeynCalc 9},
  journal       = {J. Phys. Conf. Ser.},
  volume        = {762},
  pages         = {012064},
  year          = {2016},
  eprint        = {1604.06709},
  archivePrefix = {arXiv},
  primaryClass  = {hep-ph}
}

@article{sirena2026,
  author        = {Gil, L. and L{\'o}pez Miras, J. and Moreno-S{\'a}nchez, A.},
  title         = {SIRENA---Sum-Integral REductioN Algorithm},
  journal       = {arXiv e-prints},
  eprint        = {2605.06775},
  archivePrefix = {arXiv},
  primaryClass  = {hep-ph},
  year          = {2026}
}

@article{dvornikov2014,
  author={Dvornikov, Maxim},
  title={Impossibility of the strong magnetic fields generation in an electron-positron plasma},
  journal={Phys. Rev. D}, volume={90}, pages={041702}, year={2014},
  doi={10.1103/PhysRevD.90.041702}, eprint={1405.3059},
  archivePrefix={arXiv}, primaryClass={hep-ph}
}

% \section{Exchange-bubble reduction}
% \label{app:analytic}

% The results below record the reduced static exchange-bubble coefficients at $O(k_1\chemMu)$ for graphs 6 and 7 in Fig.~\ref{fig:selected17mma}.

% The calculation is performed in the four-fermion effective theory. The loop-measure normalization is absorbed into the quantities defined below. We use
% \begin{align}
%  I_F(a,b)&= iF(a,b),
%  \nonumber\\[-2pt]
%  E^{\mu\nu}(k_1)&= \epsilon^{\mu\nu\rho}k_{1\rho}.
%  \label{eq:directDefs}
% \end{align}
% The chiral chemical-potential combination entering the reduction is
% \begin{align}
%  X&= \chemMu_{LL}
%  +4(\chemMu_{LL}-\chemMu_{LR})s_W^4
%  -4\chemMu_{LL}s_W^2
%  \nonumber\\
%  &=\chemMu_{LL}(1-4s_W^2+4s_W^4)-4\chemMu_{LR}s_W^4.
% \label{eq:Xmu}
% \end{align}
% After the spatial contractions, the trace tensor reduces to $E^{\mu\nu}(k_1)$.

\clearpage

\onecolumngrid
\begin{center}
  \textbf{\large Supplemental Material}\\[.2cm]
\end{center}
In this supplemental material, we present the sum-integral representation and non-tadpole calculations results.

\section{Sum-integral representation}
\label{app:sumint}

 For a generic fermionic momentum, the feynmann integral at finite temperature and density is
\[
\begin{aligned}
 &\sumint_{L_1,L_2}
 =
 \prod_{i=1,2}T\sum_{n_i=-\infty}^{\infty}
 \left(\frac{e^{\gamma_{\mathrm{Eul}}}\bar\Lambda^2}{4\pi}\right)^{\epsilon}
 \int\frac{\dd^{d}\bm\ell_i}{(2\pi)^d},\\
 &L_i=(i\omega_{n_i}+\chemMu_{fi},\bm\ell_i),\qquad d=3-2\epsilon.
\end{aligned}
\]
with the fermionic Matsubara frequency being $\omega_{n_i}=(2n_i+1)\pi T$.
Here $\bar\Lambda$ is the $\overline{\rm MS}$ renormalization scale. The independent fermionic loop momenta $L_1$ and $L_2$ have separate Matsubara sums and spatial integrations. For the two-loop integrals we define
\begin{equation}
\begin{aligned}
 J_{abc;rst}
 &=
 \sumint_{L_1,L_2}
 \frac{L_{10}^{\,r}L_{20}^{\,s}(L_{10}-L_{20})^{t}}
 {(L_1^2)^a(L_2^2)^b[(L_1-L_2)^2]^c},\\[-2pt]
 J_{abc;rst}
 &= I_{\rm L12}(\{a,b,c\},\{r,s,t\};\{1,1\}).
\end{aligned}
\label{eq:Jdef}
\end{equation}
The subscript $\mathrm{L12}$ labels the two-loop family built from the independent loop four-momenta $L_1$ and $L_2$, each with its own fermionic Matsubara sum and spatial integration; it is a family label, not a Lorentz index. Here $a,b,c\in\mathbb Z$ are the powers of the three quadratic denominators, while $r,s,t\in\mathbb N_0$ are the powers of the corresponding temporal components in the numerator.  A negative value of $a$, $b$, or $c$ moves the corresponding quadratic form into the numerator.  The signature $\{1,1\}$ means that both loop energies are fermionic Matsubara frequencies.  This is the two-loop specialization of the generic sum-integral notation used in Ref.~\cite{sirena2026}.

Let $\Pi^{\mu\nu}_{i,\rm L12}$ denote the fully contracted parity-odd $O(k_1\chemMu)$ contribution of an exchange-bubble graph $i$ in this basis. The representative contributions for graphs 6 and 7 are equal:
\begin{wideequations}
\begin{align}
 \Pi^{\mu\nu}_{6,\rm L12}=\Pi^{\mu\nu}_{7,\rm L12}
 ={}&-\frac{2i e^4 E^{\mu\nu}(k_1)X}{m_Z^2c_W^2s_W^2}
 \Big[
 J_{210;000}
 +2\big(
 J_{220;020}-2J_{220;110}+J_{300;000}-J_{31{-1};000}
 \nonumber\\
 &\hspace{18mm}
 +J_{310;110}-2J_{310;200}+2J_{32{-1};110}-4J_{400;200}+4J_{41{-1};200}
 \big)
 \Big].
\label{eq:L12fam67}
\end{align}
\end{wideequations}
These exchange-bubble expressions retain the $O(k_1\chemMu)$ parity-odd component required for the static photon polarization.

\section{ Direct non-tadpole two-loop calculation}
\label{app:direct}

This supplemental material records the finite coefficients obtained in the direct two-loop calculation. They provide a direct comparison of the non-tadpole Green function with the local self-energy replacement.

We define the coefficient convention
\begin{equation}
\frac{\PiTwo(0)}{\Kfac T^2}=\sum_f C_f\,\chemMu_f.
\label{eq:Cfdef}
\end{equation}
All tabulated coefficients use the reference external momentum $p=-k_1$. Table~\ref{tab:analytic} compares Eq.~(B1) of Ref.~\cite{boyarsky2012} with the direct exchange-bubble diagrams 6--17; Table~\ref{tab:numeric} gives the corresponding numerical values.

\begin{table}[!htp]
\caption{Analytic comparison of the coefficients $C_f$ in Eq.~\eqref{eq:Cfdef}, for Ref.~\cite{boyarsky2012} and the direct exchange-bubble diagrams 6--17. The $b$ row has no $Wtb$ contribution.}
\label{tab:analytic}
\centering\footnotesize
\setlength{\tabcolsep}{3.4pt}
\renewcommand{\arraystretch}{1.5}
\begin{tabular}{@{}lcc@{}}
\toprule
Coefficient & Ref. (B1) & diagrams 6--17 \\
\midrule
$C_{\nu_\alpha}$ & $-\frac{1}{27}$ & $+\frac49$ \\
$C_{\ell_\alpha}$ & $-\frac{31}{108}(1-2\cW)$ & $-\frac29(1-2\cW)$ \\
$C_u$ & $\frac{17-62\cW}{81}$ & $\frac4{81}(1+8\cW)$ \\
$C_d$ & $\frac{91+134\cW}{324}$ & $\frac2{81}(25+2\cW)$ \\
$C_b$ & $\frac{67}{324}(1+2\cW)$ & $\frac2{81}(1+2\cW)$ \\
\bottomrule
\end{tabular}
\end{table}

\begin{table}[!htp]
\caption{Numerical evaluation of Table~\ref{tab:analytic}, with the same coefficient definitions.}
\label{tab:numeric}
\centering\small
\setlength{\tabcolsep}{6pt}
\renewcommand{\arraystretch}{1.25}
\begin{tabular}{@{}lrr@{}}
\toprule
Coefficient & Ref. (B1) & diagrams 6--17 \\
\midrule
$C_{\nu_\alpha}$ & -0.03704 & +0.44444 \\
$C_{\ell_\alpha}$ & +0.02156 & +0.01669 \\
$C_u$ & -0.20159 & +0.26175 \\
$C_d$ & +0.50319 & +0.64383 \\
$C_b$ & +0.42911 & +0.05124 \\
\bottomrule
\end{tabular}
\end{table}

\end{document}